\documentclass[12pt]{article}

\usepackage[dvipdfmx]{graphicx}
\usepackage{subcaption}
\usepackage{amsmath,amssymb}
\usepackage{bm}
\usepackage{graphicx, color}
\usepackage{wrapfig}
\usepackage{cite}
\usepackage{braket}
\usepackage{float}
\begin{document}
\title{
\ \\*[-30pt]
\normalsize
{\Large \bf
Revisiting Flavor Model and Leptogenesis
\\*[20pt]}}

\author{
\centerline{
Takaaki~Nomura $^{1}$\footnote{nomura@scu.edu.cn},~
Yusuke~Shimizu $^{2}$\footnote{shimizu.yusuke@kaishi-pu.ac.jp}, and
~Towa~Takahashi $^{3}$\footnote{towa@muse.sc.niigata-u.ac.jp}}
\\*[20pt]
\centerline{
\begin{minipage}{\linewidth}
\begin{center}
$^1${\it \normalsize
College~of~Physics,~Sichuan~University,~Chengdu~610065,~China
} \\*[5pt]
$^2${\it \normalsize
Department~of~Information,~Kaishi~Professional~University, \\
Niigata~950-0916,~Japan
} \\*[5pt]
$^3${\it \normalsize 
Graduate~School~of~Science~and~Technology,~Niigata~University, \\
Niigata,~950-2181,~Japan
}
\end{center}
\end{minipage}
}
\\*[70pt]}

\date{
\centerline{\small \bf Abstract}
\begin{minipage}{0.9\linewidth}
\medskip
\medskip
\small
We revisit a supersymmetric flavor model based on the symmetries 
$SU(2)_L \times A_4 \times Z_3 \times U(1)_R$, 
which extends the original Altarelli and Feruglio construction by introducing flavon and driving superfields responsible for the spontaneous breaking of the flavor symmetry in order to obtain non-zero reactor angle. 
The vacuum alignments of flavon fields are achieved through the minimization of the scalar potential derived from the superpotential. This setup leads to specific mass matrices for the charged leptons and neutrinos that are consistent with current experimental data, including the measured values of the lepton mixing angles and neutrino mass squared differences.
We investigate whether the model can simultaneously accommodate successful thermal leptogenesis. 
In particular, we analyze the CP asymmetry generated in the decay of heavy Majorana neutrinos, the resulting lepton asymmetry, and its conversion to the baryon asymmetry through the electroweak sphalerons. 
However the CP asymmetry is zero, since the Dirac neutrino mass matrix is simple texture in the leading order for our model. Then we consider the next-to-leading order in Yukawa interactions of the Dirac neutrinos. Therefore, we can realize the baryon asymmetry of the universe at the present universe. 
By numerically scanning the parameter space, we identify the regions consistent with both neutrino oscillation data and the observed baryon asymmetry. 
In the specific case such that one of the couplings for the right-handed Majorana neutrinos is real parameter, the predicted lightest neutrino mass is at least $5$ meV and $15$ meV for the normal and inverted neutrino mass hierarchies, respectively. 
In addition, the range of the Majorana phases may be tested in future experiments.
\end{minipage}
}

\begin{titlepage}
\maketitle
\thispagestyle{empty}
\end{titlepage}
\newpage
\section{Introduction}

The standard model (SM) of particle physics has successfully described a wide range of experimental phenomena, culminating in the discovery of the Higgs boson. 
However, There are several crucial mysteries, the origin of neutrino masses and mixing angles, the baryon asymmetry of the universe (BAU), and the flavor structure of the fermion sector. 
In particular, the discovery of the neutrino oscillations has established that neutrinos are massive and the lepton mixing angles are different from that of the quark sector. 
Then, we need the new physics beyond the SM.

The observed pattern of lepton mixing, characterized by large mixing angles and CP violation in the lepton sector, poses a significant theoretical challenge. 
A promising direction to address this issue involves the introduction of discrete flavor symmetries~\cite{Altarelli:2010gt}-\cite{Kobayashi:2022moq}, which can generate specific textures of mass matrices. 
Among these, the non-Abelian discrete group $A_4$ has emerged as a effective choice. 
The original model proposed by Altarelli and Feruglio (AF) employs the $A_4$ symmetry to realize the so-called tri-bimaximal (TBM) mixing pattern~\cite{Harrison:2002er,Harrison:2002kp}, which was in good agreement with early neutrino data in Ref.~\cite{Altarelli:2005yp,Altarelli:2005yx}. 
Since the precise measurements of the reactor angle $\theta_{13}$~\cite{DayaBay:2012fng,RENO:2012mkc} has shown a deviation from TBM, we should consider the deviation or breaking the TBM~\cite{Xing:2002sw}-\cite{King:2013eh} or other patterns of the lepton mixing angles~\cite{King:2012vj}-\cite{Shimizu:2012ry}. 
In particular modified versions of the AF model still offer an attractive framework for explaining the observed mixing angles~\cite{Muramatsu:2016bda}-\cite{Nomura:2024abu}.

In parallel, the observed baryon asymmetry of the universe, quantified by the baryon-to-photon ratio cannot be explained within the SM due to insufficient sources of CP violation and the lack of a mechanism for out-of-equilibrium baryon number violation at the required scale. 
The leptogenesis~\cite{Fukugita:1986hr} is a compelling scenario that addresses this issue by generating a lepton asymmetry in the early universe, which is partially converted into a baryon asymmetry via sphaleron processes. In the type-I seesaw framework~\cite{Minkowski:1977sc}-\cite{Schechter:1980gr}, the heavy right-handed Majorana neutrinos decay out of equilibrium, leading to CP asymmetries~\cite{Covi:1996wh}-
\cite{Giudice:2003jh} that can account for the observed BAU.

The connection between flavor models and leptogenesis is particularly intriguing. The flavor structure imposed by a discrete symmetry such as $A_4$ constrains the form of the neutrino Yukawa couplings and the Majorana mass matrix of right-handed neutrinos. These constraints not only determine the low-energy neutrino observables but also affect the dynamics of leptogenesis, including the CP asymmetry~\cite{Endoh:2002wm,Pascoli:2006ci}. Consequently, it becomes possible to establish a direct link between the flavor symmetry breaking pattern and the generation of the baryon asymmetry.

In this paper, we revisit a supersymmetric flavor model based on the 
$SU(2)_L \times A_4 \times Z_3 \times U(1)_R$ symmetries, 
which extends the original AF construction in order to obtain the non-zero reactor angle by introducing flavon and driving superfields responsible for the spontaneous breaking of the flavor symmetry. 
The vacuum alignment of flavon fields is achieved through the minimization of the scalar potential derived from the superpotential. 
This setup leads to specific mass matrices for the charged leptons and neutrinos that are consistent with current experimental data, including the measured values of the mixing angles and mass squared differences.
We investigate whether the model can simultaneously accommodate successful thermal leptogenesis. 
In particular, we analyze the CP asymmetries generated in the decay of heavy Majorana neutrinos, the resulting lepton asymmetry, and its conversion to the baryon asymmetry through electroweak sphalerons. 
However the CP asymmetry is zero, since the Dirac neutrino mass matrix is simple texture in the leading order for our model. Then we consider the next-to-leading order (NLO) in Yukawa interactions of the Dirac neutrinos.
We also study the impact of flavor effects and the role of specific vacuum expectation value (VEV) alignments in determining the relevant mass parameters and Yukawa couplings. By numerically scanning the parameter space, we identify the regions consistent with both neutrino oscillation data and the observed baryon asymmetry.

This paper is organized as follows. In Section~\ref{sec:Model}, we outline the structure of the model, including the field content, charge assignments, and the construction of the superpotential. 
Section~\ref{sec:Leptogenesis} is devoted to the study of leptogenesis, where we consider the CP asymmetry in our model.
In Section~\ref{sec:Numerical}, we present the resulting mass matrices and their implications for neutrino phenomenology. 
We summarize our results and conclude in Section~\ref{sec:Summary}. 
The multiplication rule of $A_4$ is included in the Appendix~\ref{sec:multiplication-rule}.

\section{Model}
\label{sec:Model}
In this section, we briefly introduce an $SU(2)_L\times A_4\times Z_3\times U(1)_R$ 
model~\cite{Shimizu:2011xg},\cite{Muramatsu:2016bda}-\cite{Muramatsu:2017xmn}. 
Three generations of the left-handed lepton doublet superfields 
$\Phi _\ell =(\Phi _{\ell 1},\Phi _{\ell 2},\Phi _{\ell 3})$ are assigned to the triplet as $\bf 3$ for $A_4$ symmetry. 
The right-handed charged lepton superfields $\Phi _{e_R^c}$, $\Phi _{\mu _R^c}$, and $\Phi _{\tau _R^c}$ are 
assigned to the singlets as $\bf 1$, $\bf 1''$, and $\bf 1'$ for $A_4$ symmetry, respectively. 
The Higgs doublet superfields are defined as $\Phi _u$ and $\Phi _d$ which are assigned to the singlet as 
$\bf 1$ for $A_4$ symmetry. 
We introduce three generations of the right-handed neutrino superfields 
$\Phi _N=(\Phi _{N1},\Phi _{N2},\Phi _{N3})$ 
which are assigned to the triplet as $\bf 3$ for $A_4$ symmetry. 
We also introduce superfields $\Phi _T=(\Phi _{T1},\Phi _{T2},\Phi _{T3})$, $\Phi _S=(\Phi _{S1},\Phi _{S2},\Phi _{S3})$, 
$\Phi _\xi $, and $\Phi _{\xi '}$ 
which are so-called ``flavon superfields'' and assigned to the triplets as $\bf 3$ 
and singlets as $\bf 1$, and $\bf 1'$ for $A_4$ symmetry, respectively. 
In order to obtain the relevant couplings, we add the $Z_3$ symmetry. 
We also add the $U(1)_R$ symmetry so that we can generate the VEVs and VEV alignments 
through the $F$-terms by coupling flavon superfields so-called 
``driving superfields'' $\Phi _0^T=(\Phi _{01}^T,\Phi _{02}^T,\Phi _{03}^T)$ , 
$\Phi _0^S=(\Phi _{01}^S,\Phi _{02}^S,\Phi _{03}^S)$, $\Phi _0^\xi $ 
which are assigned to triplets as $\bf 3$ and trivial singlet as $\bf 1$ for $A_4$ symmetry 
and carry the $R$ charge $+2$ under $U(1)_R$ symmmetry.
The charge assignments of $SU(2)_L$, $A_4$, $Z_3$, and $U(1)_R$ are summarized in Tab.~\ref{tab:model}.
\begin{table}[h]
  \centering
  \begin{tabular}{|c||ccccc|ccccc|ccc|}
    \hline 
    \rule[14pt]{0pt}{0pt}
    & $\Phi _\ell $ & $\Phi _{e_R^c}$ & $\Phi _{\mu_R^c}$ & 
    $\Phi _{\tau_R^c}$ & $\Phi _N$ & 
    $\Phi _{u,d}$ & $\Phi _T$ & $\Phi _S$ & $\Phi _\xi $ & $\Phi _{\xi '}$ & 
    $\Phi _0^T$ & $\Phi _0^S$ & $\Phi _0^\xi $ 
    \\ \hline 
    \rule[14pt]{0pt}{0pt}
    $SU(2)_L$ & $2$ & $1$ & $1$ & $1$ & $1$ & $2$ & $1$ & $1$ & $1$ & $1$ & $1$ & $1$ & $1$
    \\
    $A_4$ & ${\bf 3}$ & ${\bf 1}$ & ${\bf 1''}$ & ${\bf 1'}$ & ${\bf 3}$ & ${\bf 1}$ & ${\bf 3}$ & ${\bf 3}$ & 
    ${\bf 1}$ & ${\bf 1'}$ & ${\bf 3}$ & ${\bf 3}$ & ${\bf 1}$\\
    $Z_3$ & $\omega $ & $\omega ^2$ & $\omega^2$ & $\omega ^2$ & $\omega ^2$ & 
    $1$ & $1$ & $\omega ^2$ & $\omega ^2$ & $\omega ^2$ & $1$ & $\omega ^2$ & $\omega ^2$\\
    $U(1)_R$ & $1$ & $1$ & $1$ & $1$ & $1$ & $0$ & $0$ & $0$ & $0$ & $0$ & $2$ & $2$ & $2$ \\
    \hline
  \end{tabular}
  \caption{The charge assignments of $SU(2)_L\times A_4\times Z_3\times U(1)_R$ symmetry in our model.}
  \label{tab:model}
\end{table}
In these setup, we can now write down the superpotential for respecting 
$SU(2)_L\times A_4\times Z_3\times U(1)_R$ symmetry at the leading order in terms of 
the $A_4$ cut-off scale $\Lambda $ as 
\begin{align}
w&=w_Y+w_d, \nonumber \\
w_Y&=w_{\ell }+w_D+w_N, \nonumber \\
w_{\ell }&=y_e\Phi _T\Phi _\ell \Phi _{e_R^c}\Phi _d/\Lambda + y_\mu \Phi _T\Phi _\ell \Phi _{\mu _R^c}\Phi _d/\Lambda + y_\tau \Phi _T\Phi _\ell \Phi _{\tau _R^c}\Phi _d/\Lambda ,\nonumber \\
w_D&=y_D\Phi _\ell \Phi _N\Phi _u, \nonumber \\
w_N&=y_{\Phi _S}\Phi _N\Phi _N \Phi _S+y_\xi \Phi _N\Phi _N \Phi _\xi +y_{\Phi _{\xi '}}\Phi _N\Phi _N \Phi _{\xi '}, \nonumber \\
w_d&=w_d^T+w_d^S, \nonumber \\
w_d^T&=-M\Phi _0^T\Phi _T+g\Phi _0^T\Phi _T\Phi _T, \nonumber \\
w_d^S&=g_1\Phi _0^S\Phi _S\Phi _S+g_2\Phi _0^S\Phi _S\Phi _\xi +g_2'\Phi _0^S\Phi _S\Phi _{\xi '}+g_3\Phi _0^\xi \Phi _{S}\Phi _{S}-g_4\Phi _0^\xi \Phi _\xi \Phi _\xi ,
\label{eq:superpotential}
\end{align}
where $\Phi _i$'s are chiral superfields $\Phi _i=\phi _i+\sqrt{2}\theta \psi _i+\theta \theta F_i$, $y_j$'s are Yukawa couplings, $M$ is mass paremeter for the flavon and driving superfields $\Phi _T$ and $\Phi _0^T$, and $g$ and $g_k$'s are trilinear couplings for flavon and driving superfields. 
The superpotential in Eq.~\eqref{eq:superpotential} related to the Yukawa interactions is rewritten as
\begin{align}
w_{\ell }&=y_e(\Phi _{T1}\Phi _{\ell 1}+\Phi _{T2}\Phi _{\ell 3}+\Phi _{T3}\Phi _{\ell 2})\Phi _{e_R^c}\Phi _d/\Lambda \nonumber \\
&+y_\mu (\Phi _{T3}\Phi _{\ell 3}+\Phi _{T1}\Phi _{\ell 2}+\Phi _{T2}\Phi _{\ell 1})\Phi _{\mu _R^c}\Phi _d/\Lambda \nonumber \\
&+y_\tau (\Phi _{T2}\Phi _{\ell 2}+\Phi _{T3}\Phi _{\ell 1}+\Phi _{T1}\Phi _{\ell 3})\Phi _{\tau _R^c}\Phi _d/\Lambda , \nonumber \\
w_D&=y_D(\Phi _{\ell 1}\Phi _{N1}+\Phi _{\ell 2}\Phi _{N3}+\Phi _{\ell 3}\Phi _{N2})\Phi _u, \nonumber \\
w_N&=\frac{1}{3}y_{\Phi _S}\big [(2\Phi_{N1}\Phi_{N1}-\Phi_{N2}\Phi_{N3}-\Phi_{N3}\Phi_{N2})\Phi _{S1}+(2\Phi_{N2}\Phi_{N2}-\Phi_{N3}\Phi_{N1}-\Phi_{N1}\Phi_{N3})\Phi _{S2} \nonumber \\
&+(2\Phi_{N3}\Phi_{N3}-\Phi_{N1}\Phi_{N2}-\Phi_{N2}\Phi_{N1})\Phi _{S3}\big ] \nonumber \\
&+y_\xi (\Phi _{N1}\Phi_{N1}+\Phi_{N2}\Phi_{N3}+\Phi_{N3}\Phi_{N2})\Phi_\xi 
+y_{\xi '}(\Phi _{N2}\Phi_{N2}+\Phi_{N3}\Phi_{N1}+\Phi_{N1}\Phi_{N3})\Phi_{\xi '}.
\label{eq:Yukawa}
\end{align}
On the other hand, the superpotential in Eq.~\eqref{eq:superpotential} which are related to the scalar potential with driving superfields is rewritten as 
\begin{align}
w_d^T&=-M(\Phi _{01}^T\Phi _{T1}+\Phi _{02}^T\Phi _{T3}+\Phi _{03}^T\Phi _{T2}) \nonumber \\
&+\frac{2}{3}g\left [\Phi _{01}^T(\Phi _{T1}\Phi _{T1}-\Phi _{T2}\Phi _{T3})+\Phi _{02}^T(\Phi _{T2}\Phi _{T2}-\Phi _{T3}\Phi _{T1})+\Phi _{03}^T(\Phi _{T3}\Phi _{T3}-\Phi _{T1}\Phi _{T2})\right ], \nonumber \\
w_d^S&=\frac{1}{3}g_1\big [\Phi _{01}^S(2\Phi _{S1}\Phi _{S1}-\Phi _{S2}\Phi _{S3}-\Phi _{S3}\Phi _{S2})+\Phi _{02}^S(2\Phi _{S2}\Phi _{S2}-\Phi _{S3}\Phi _{S1}-\Phi _{S1}\Phi _{S3}) \nonumber \\
&+\Phi _{03}^S(2\Phi _{S3}\Phi _{S3}-\Phi _{S1}\Phi _{S2}-\Phi _{S2}\Phi _{S1})\big ] \nonumber \\
&+g_2(\Phi _{01}^S\Phi _{S1}+\Phi _{02}^S\Phi _{S3}+\Phi _{03}^S\Phi _{S2})\Phi _\xi 
+g_2'(\Phi _{02}^S\Phi _{S2}+\Phi _{03}^S\Phi _{S1}+\Phi _{01}^S\Phi _{S3})\Phi _{\xi '} \nonumber \\
&+g_3\Phi _0^\xi (\Phi _{S1}\Phi _{S1}+\Phi _{S2}\Phi _{S3}+\Phi _{S3}\Phi _{S2})-g_4\Phi _0^\xi \Phi _\xi \Phi _\xi .
\end{align}
Then, the Lagrangian $\mathcal {L}$ for our model is given as
\begin{align}
\mathcal{L}&=\mathcal{L}_Y+\mathcal{L}_d -V\nonumber \\
\mathcal{L}_Y&=\int d^2\theta w_Y+\int d^2\bar \theta \bar w_Y, \nonumber \\
\mathcal{L}_d&=\int d^2\theta w_d+\int d^2\bar \theta \bar w_d, \nonumber \\
V&=V_Y+V_d .
\end{align}
where the scalar potential $V_Y$ and $V_d$ are obtained from $w_Y$ and $w_d$, respectively. 
In this paper, we focus on the phenomenology of the SM. Therefore, we consider only $\mathcal{L}_Y$ and $V_d$. 
The scalar potential $V_d$ is obtained as 
\begin{align}
V_d&=V_T+V_S, 
\end{align}
where 
\begin{align}
V_T&=\sum _X\left |\frac{\partial w_d^T}{\partial X}\right |^2 \nonumber \\
&=\left | -M\phi _{01}^T+\frac{2}{3}g\left (2\phi _{01}^T\phi _{T1}-\phi _{02}^T\phi _{T3}-\phi _{03}^T\phi _{T2}\right )\right |^2 \nonumber \\
&+\left | -M\phi _{03}^T+\frac{2}{3}g\left (-\phi _{01}^T\phi _{T3}+2\phi _{02}^T\phi _{T2}-\phi _{03}^T\phi _{T1}\right )\right |^2 \nonumber \\
&+\left | -M\phi _{02}^T+\frac{2}{3}g\left (-\phi _{01}^T\phi _{T2}-\phi _{02}^T\phi _{T1}+2\phi _{03}^T\phi _{T3}\right )\right |^2 \nonumber \\
&+\left | -M\phi _{T1}+\frac{2}{3}g\left (\phi _{T1}\phi _{T1}-\phi _{T2}\phi _{T3}\right )\right |^2 \nonumber \\
&+\left | -M\phi _{T3}+\frac{2}{3}g\left (\phi _{T2}\phi _{T2}-\phi _{T3}\phi _{T1}\right )\right |^2 \nonumber \\
&+\left | -M\phi _{T2}+\frac{2}{3}g\left (\phi _{T3}\phi _{T3}-\phi _{T1}\phi _{T2}\right )\right |^2 ,
\end{align}
and 
\begin{align}
V_S&=\sum _Y\left |\frac{\partial w_d^S}{\partial Y}\right |^2 \nonumber \\
&=\left |\frac{2}{3}g_1\left (2\phi _{01}^S\phi _{S1}-\phi _{02}^S\phi _{S3}-\phi _{03}^S\phi _{S2}\right )
+g_2\phi _{01}^S\phi _\xi +g'_2\phi _{03}^S\phi _{\xi '}+2g_3\phi _0^\xi \phi _{S1}\right |^2 \nonumber \\
&+\left |\frac{2}{3}g_1\left (-\phi _{01}^S\phi _{S3}+2\phi _{02}^S\phi _{S2}-\phi _{03}^S\phi _{S1}\right )
+g_2\phi _{03}^S\phi _\xi +g'_2\phi _{02}^S\phi _{\xi '}+2g_3\phi _0^\xi \phi _{S3}\right |^2 \nonumber \\
&+\left |\frac{2}{3}g_1\left (-\phi _{01}^S\phi _{S2}-\phi _{02}^S\phi _{S1}+2\phi _{03}^S\phi _{S3}\right )
+g_2\phi _{02}^S\phi _\xi +g'_2\phi _{01}^S\phi _{\xi '}+2g_3\phi _0^\xi \phi _{S2}\right |^2 \nonumber \\
&+\left |\frac{2}{3}g_1\left (\phi _{S1}\phi _{S1}-\phi _{S2}\phi _{S3}\right )
+g_2\phi _{S1}\phi _\xi +g'_2\phi _{S3}\phi _{\xi '}\right |^2 \nonumber \\
&+\left |\frac{2}{3}g_1\left (\phi _{S2}\phi _{S2}-\phi _{S3}\phi _{S1}\right )
+g_2\phi _{S3}\phi _\xi +g'_2\phi _{S2}\phi _{\xi '}\right |^2 \nonumber \\
&+\left |\frac{2}{3}g_1\left (\phi _{S3}\phi _{S3}-\phi _{S1}\phi _{S2}\right )
+g_2\phi _{S2}\phi _\xi +g'_2\phi _{S1}\phi _{\xi '}\right |^2 \nonumber \\
&+\left |g_2\left (\phi _{01}^S\phi _{S1}+\phi _{02}^S\phi _{S3}+\phi _{03}^S\phi _{S2}\right )-2g_4\phi _0^\xi \phi _\xi \right |^2 \nonumber \\
&+\left |g'_2\left (\phi _{02}^S\phi _{S2}+\phi _{03}^S\phi _{S1}+\phi _{01}^S\phi _{S3}\right ) \right |^2 \nonumber \\
&+\left |g_3\left (\phi _{S1}\phi _{S1}+2\phi _{S2}\phi _{S3}\right )-g_4\phi _\xi \phi _\xi \right |^2 .
\end{align}
The sum for $X$, $Y$ runs over all the scalar fields: 
$X=\phi _{T1}, \phi _{T2}, \phi _{T3}, \phi _{01}^T, \phi _{02}^T$, and $\phi _{03}^T$, 
$Y=\phi _{S1}, \phi _{S2}, \phi _{S3}, \phi _{01}^S, \phi _{02}^S, \phi _{03}^S, \phi _\xi ,\phi _{\xi '}$, and $\phi _0^\xi $.
The VEV alignments of $\phi_T$ and $\phi _0^T$ are obtained from the potential minimum condition, $V_T =0$, as in Ref.~\cite{Morozumi:2017rrg};
\begin{equation}
\langle \phi_T \rangle = v_T (1, 0, 0), \qquad  v_T = \frac{3M}{2g}, \qquad \langle \phi _0^T\rangle =(0,0,0).
\label{eq:VT_VEV_alignments}
\end{equation} 
On the other hand, the VEV alignments of $\phi _S$ and $\phi _0^S$, and VEVs of $\phi _\xi $, $\phi _{\xi '}$, and $\phi _0^\xi $ are obtained from the potential minimum condition, 
$V_S=0$, as in Ref.~\cite{Morozumi:2017rrg};
\begin{equation}
\langle \phi _S\rangle = v_S(1,1,1),\quad v_S=\sqrt{\frac{g_4}{3g_3}}u,\quad \langle \phi _0^S\rangle =(0,0,0),
\quad \langle \phi _{\xi '}\rangle =u'=\frac{g_2}{g_2'}u,\quad \langle \phi _\xi \rangle =u,\quad \langle \phi _0^\xi \rangle =0.
\label{eq:VS_VEV_alignments}
\end{equation}
By using the VEVs and VEV alignments in Eqs.~\eqref{eq:VT_VEV_alignments} and \eqref{eq:VS_VEV_alignments}, 
we briefly explain the mass matrices for charged leptons and neutrinos after taking VEVs $\langle \phi _u\rangle =v_u$ and 
$\langle \phi _d\rangle =v_d$ for two Higgs doublets.

The mass matrix of the charged leptons $M_\ell $ is written in Eq.~\eqref{eq:Yukawa},
\begin{equation}
M_\ell =\frac{v_dv_T}{\Lambda }
\begin{pmatrix}
y_e & 0 & 0 \\
0 & y_\mu & 0 \\
0 & 0 & y_\tau 
\end{pmatrix}.
\label{eq:charged-lepton-mass-matrix}
\end{equation}
The mass matrix of the Dirac neutrinos $M_D$ is written in Eq.~\eqref{eq:Yukawa},
\begin{equation}
M_D =y_Dv_u
\begin{pmatrix}
1 & 0 & 0 \\
0 & 0 & 1 \\
0 & 1 & 0 
\end{pmatrix}.
\label{eq:Dirac-neutrino-mass-matrix}
\end{equation}
On the other hand, the mass matrix of the right-handed Majorana neutrinos $M_R$ is written in Eq.~\eqref{eq:Yukawa},
\begin{equation}
M_R =\frac{1}{3}y_{\phi _S}v_S
\begin{pmatrix}
2 & -1 & -1 \\
-1 & 2 & -1 \\
-1 & -1 & 2
\end{pmatrix}+y_\xi u
\begin{pmatrix}
1 & 0 & 0 \\
0 & 0 & 1 \\
0 & 1 & 0
\end{pmatrix}+y_{\xi'} u'
\begin{pmatrix}
0 & 0 & 1 \\
0 & 1 & 0 \\
1 & 0 & 0
\end{pmatrix}.
\label{eq:right-handed-Majorana-neutrino-mass-matrix}
\end{equation}
By using the type-I seesaw mechanism $M_\nu =-M_DM_R^{-1}M_D^T$ in Refs.~\cite{Minkowski:1977sc}-\cite{Schechter:1980gr}, the mass matrix of the left-handed Majorana neutrinos is given as
\begin{equation}
M_\nu =a
\begin{pmatrix}
1 & 0 & 0 \\
0 & 1 & 0 \\
0 & 0 & 1
\end{pmatrix}+b
\begin{pmatrix}
1 & 1 & 1 \\
1 & 1 & 1 \\
1 & 1 & 1 
\end{pmatrix}+c
\begin{pmatrix}
1 & 0 & 0 \\
0 & 0 & 1 \\
0 & 1 & 0
\end{pmatrix}+d
\begin{pmatrix}
0 & 0 & 1 \\
0 & 1 & 0 \\
1 & 0 & 0
\end{pmatrix},
\label{eq:left-handed-Majorana-neutrino-mass-matrix}
\end{equation}
where $a$, $b$, $c$, and $d$ are combinations for the Yukawa couplings in Eq.~\eqref{eq:Yukawa} and VEV and VEV alignments in Eqs.~\eqref{eq:VT_VEV_alignments} and \eqref{eq:VS_VEV_alignments}. The more details are shown in Ref.~~\cite{Morozumi:2017rrg}.
We show the numerical analyses after diagonalizing the mass matrices in Eqs.~\eqref{eq:charged-lepton-mass-matrix} and \eqref{eq:left-handed-Majorana-neutrino-mass-matrix} after discussing leptogenesis in the model below. 
%
%


\section{Leptogenesis in flavor model}
\label{sec:Leptogenesis}
In this section, we discuss the leptogenesis in our model. 
The BAU at the present universe is measured very precisely by the cosmic microwave background
radiation in Ref.~\cite{Planck:2018vyg} as 
\begin{equation}
Y_B=\frac{n_B}{s}=(0.852-0.888)\times 10^{-10}, 
\end{equation}
at $3\sigma $ confidence level, where $Y_B$ is defined by the ratio between the number density of baryon asymmetry $n_B$ and the entropy density $s$. 
One of the attractive scenarios for baryogenesis is the canonical leptogenesis scenario~\cite{Fukugita:1986hr} in which the decays of right-handed Majorana neutrinos can generate the lepton asymmetry that is partially converted into the baryon asymmetry via the sphaleron process~\cite{Kuzmin:1985mm}.
The CP asymmetry parameter is given in Refs.~\cite{Covi:1996wh}-\cite{Giudice:2003jh}:
\begin{align}
\epsilon _I&=\frac{\Gamma \left (N_I\to \ell +\bar {h_u}\right )-\Gamma \left (N_I\to \bar \ell +h_u \right )}{\Gamma \left (N_I\to \ell +\bar {h_u}\right )+\Gamma \left (N_I\to \bar \ell +h_u \right )} \nonumber \\
&=-\frac{1}{8\pi }\sum _{J\neq I}\frac{\text{Im}\left [\left \{ \left (Y_D^\dagger Y_D\right )_{JI}\right \} ^2\right ]}{\left (Y_D^\dagger Y_D\right )_{II}}\left [f^V\left (\frac{M_J^2}{M_I^2}\right )+f^S\left (\frac{M_J^2}{M_I^2}\right )\right ],
\end{align}
where $N_I~(I=1, 2, 3)$ are the right-handed Majorana neutrinos, $\ell $ is lepton doublet, and 
 $f^V(x)$ and $f^S(x)$ are the contributions from vertex and self-energy corrections, respectively.
In the case of the SM with right-handed Majorana neutrinos, they are given as
\begin{equation}
f^V(x)=\sqrt{x}\left [(x+1)\ln \left (1+\frac{1}{x}\right )-1\right ],\quad f^S(x)=\frac{\sqrt{x}}{x-1}.
\end{equation}
Then, the CP asymmetry is proportional to the imaginary part of the Yukawa couplings as
\begin{equation}
\epsilon _I \propto \sum _{J\neq I}\text{Im}\left [\left \{ \left (Y_D^\dagger Y_D\right )_{JI}\right \} ^2\right ], 
\end{equation}
where $N_I~(I=1, 2, 3)$ are the right-handed Majorana neutrinos and $Y_D^{\dagger }Y_D$ is given by the Dirac neutrino mass matrix $M_D$ in the real diagonal base for the right-handed Majorana neutrino mass matrix $M_R$.

The Dirac neutrino mass matrix is given in Eq.~\eqref{eq:Dirac-neutrino-mass-matrix}.
Then, the Dirac Yukawa matrix in the real diagonal base for the right-handed Majorana neutrino mass matrix $Y_D^\text{L}$ is written as 
\begin{equation}
Y_D^\text{L}=
y_D
\begin{pmatrix}
1 & 0 & 0 \\
0 & 0 & 1 \\
0 & 1 & 0
\end{pmatrix}U_R^TP_R^T ,
\end{equation}
which comes from the leading order of the Dirac neutrino Yukawa interactions.
The $Y_D^{\text{L}\dagger }Y_D^\text{L}$ is obtained as 
\begin{align}
Y_D^{\text{L}\dagger }Y_D^\text{L}&=|y_D|^2P_R^*U_R^*
\begin{pmatrix}
1 & 0 & 0 \\
0 & 0 & 1 \\
0 & 1 & 0
\end{pmatrix}
\begin{pmatrix}
1 & 0 & 0 \\
0 & 0 & 1 \\
0 & 1 & 0
\end{pmatrix}
U_R^TP_R^T \nonumber \\
&=|y_D|^2
\begin{pmatrix}
1 & 0 & 0 \\
0 & 1 & 0 \\
0 & 0 & 1
\end{pmatrix}.
\end{align}
Therefore the CP asymmetry parameter is zero. In order to obtain the non-zero CP asymmetry, 
we consider the NLO for the Yukawa interactions of the Dirac neutrinos. 
The superpotential for the NLO in the Yukawa interactions of the Dirac neutrinos is written as
\begin{align}
w_D^\text{NL}&=y_D^\text{NL}\Phi _\ell \Phi _N\Phi _u\Phi _T/\Lambda \nonumber \\
&=\frac{1}{3}y_D^\text{NLS}\big [(2\Phi_{\ell 1}\Phi_{N1}-\Phi_{\ell 2}\Phi_{N3}-\Phi_{\ell 3}\Phi_{N2})\Phi _{T1}+(2\Phi_{\ell 2}\Phi_{N2}-\Phi_{\ell 3}\Phi_{N1}-\Phi_{\ell 1}\Phi_{N3})\Phi _{T2} \nonumber \\
&+(2\Phi_{\ell 3}\Phi_{N3}-\Phi_{\ell 1}\Phi_{N2}-\Phi_{\ell 2}\Phi_{N1})\Phi _{T3}\big ]\Phi _u/\Lambda \nonumber \\
&+\frac{1}{2}y_D^\text{NLA}\big [(\Phi_{\ell 2}\Phi_{N3}-\Phi_{\ell 3}\Phi_{N2})\Phi _{T1}+(\Phi_{\ell 3}\Phi_{N1}-\Phi_{\ell 1}\Phi_{N3})\Phi _{T2}+(\Phi_{\ell 1}\Phi_{N2}-\Phi_{\ell 2}\Phi_{N1})\Phi _{T3}\big ]\Phi _u/\Lambda .
\end{align}
By taking VEVs and VEV alignments in Eqs.~\eqref{eq:VT_VEV_alignments} and \eqref{eq:VS_VEV_alignments}, the Dirac Yukawa matrix including the NLO $Y_D^\text{L+NL}$ is obtained as
\begin{equation}
Y_D^\text{L+NL}=y_D
\begin{pmatrix}
1 & 0 & 0 \\
0 & 0 & 1 \\
0 & 1 & 0
\end{pmatrix}+\frac{1}{3}y_D^\text{NLS}\frac{v_T}{\Lambda }
\begin{pmatrix}
2 & 0 & 0 \\
0 & 0 & -1 \\
0 & -1 & 0
\end{pmatrix}+\frac{1}{2}y_D^\text{NLA}\frac{v_T}{\Lambda }
\begin{pmatrix}
0 & 0 & 0 \\
0 & 0 & 1 \\
0 & -1 & 0
\end{pmatrix},
\label{eq:Dirac-Yukawa-matrix-NL}
\end{equation}
where $y_D^\text{NLS}$ and $y_D^\text{NLD}$ are the Dirac Yukawa couplings in the NLO.
By using the formulations in Sections~\ref{sec:Model} and \ref{sec:Leptogenesis}, we will discuss the numerical analyses in the next section.


\section{Numerical analysis}
\label{sec:Numerical}
In this section, we show the numerical analyses such as lepton flavor mixing angles, Dirac CP phase, Majorana phases, and the effective mass for the neutrino less double beta (0$\nu \beta \beta$) decay. Here we take into account for the CP asymmetry numerically. 

First of all, we introduce the PDG parametrization~\cite{ParticleDataGroup:2024cfk} of the PMNS matrix~\cite{Maki:1962mu,Pontecorvo:1967fh}: 
\begin{align}
  U_{PMNS}^{pdg} = 
      \begin{pmatrix}
        1 & 0 & 0  &  \\
        0 & c_{23} & s_{23} \\
        0 & -s_{23} & c_{23} 
      \end{pmatrix}
      \begin{pmatrix}
        c_{13} & 0 & s_{13}e^{-i \delta_{CP}}\\
        0 & 1 & 0 \\
        -s_{13}e^{i \delta_{CP}} & 0 & c_{13}
      \end{pmatrix}
      \begin{pmatrix}
        c_{12} & s_{12} & 0\\
        -s_{12} & c_{12} & 0\\
        0 & 0 & 1
      \end{pmatrix}
      \begin{pmatrix}
        e^{i \eta_1} &0&0\\
        0& e^{i \eta_2} & 0\\
        0&0&1\\
      \end{pmatrix}
\end{align}
where $s_{ij}$ and $c_{ij}$ represent lepton mixing angles $\sin \theta_{ij}$ and $\cos \theta_{ij}$, respectively.
The lepton mixing angles are written as follows:
\begin{align}
  \sin^2 \theta_{12} = \frac{|U_{e2}|^2}{1 - |U_{e2}|^2},
  ~~~~~
  \sin^2 \theta_{23} = \frac{|U_{\mu 3}|^2}{1 - |U_{e3}|^2},
  ~~~~~
  \sin^2 \theta_{13} = |U_{e3}|^2,
\end{align}
where $U_{\alpha i}$ represent the PMNS matrix elements.
The Dirac CP violating phase $\delta_{CP}$ can be obteined by the Jarlskog invariant:
\begin{align}
  \sin \delta_{CP} = \frac{J_{CP}}{s_{23}c_{23}s_{12}c_{12}s_{13}c_{12}^2},\\
  J_{CP} = \text{Im} [U_{e1}U_{\mu 2}U_{\mu 1}^*U_{e2}^*].
\end{align}
The $\delta_{CP}$ is also determined by one of the absolute values for PMNS matrix elements:
\begin{align}
  |U_{\tau 1}|^2 = s_{12}^2 s_{23}^2 + c_{12}^2 c_{23}^2 s_{13}^2 - 2 s_{12} s_{23} c_{12} c_{23} s_{13} \cos \delta_{CP}
\end{align}
Then, we can determine the $\delta_{CP}$.

\begin{figure}[h]
  \centering
        \includegraphics[width=0.48\columnwidth]{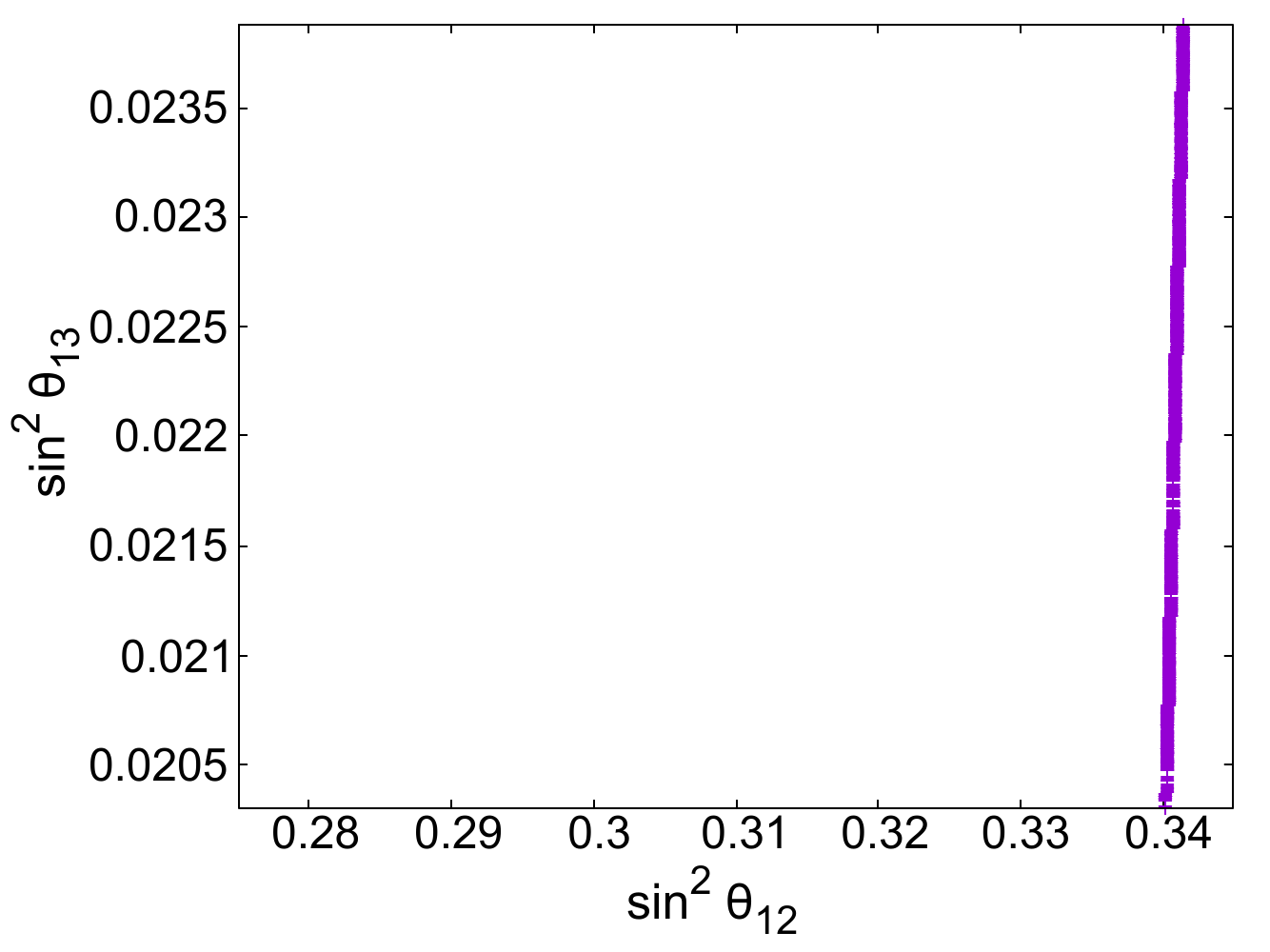}
    \caption{The allowed regions of the lepton mixing anlges and Dirac CP phase. Horizontal axis:$\sin \theta_{12}$, Vertical axis:$\sin \theta_{13}$.}
\label{fig:mixing-angles}
\end{figure}

\begin{figure}[h]
  \centering
    \begin{minipage}[b]{0.48\linewidth}
        \includegraphics[width=0.9\columnwidth]{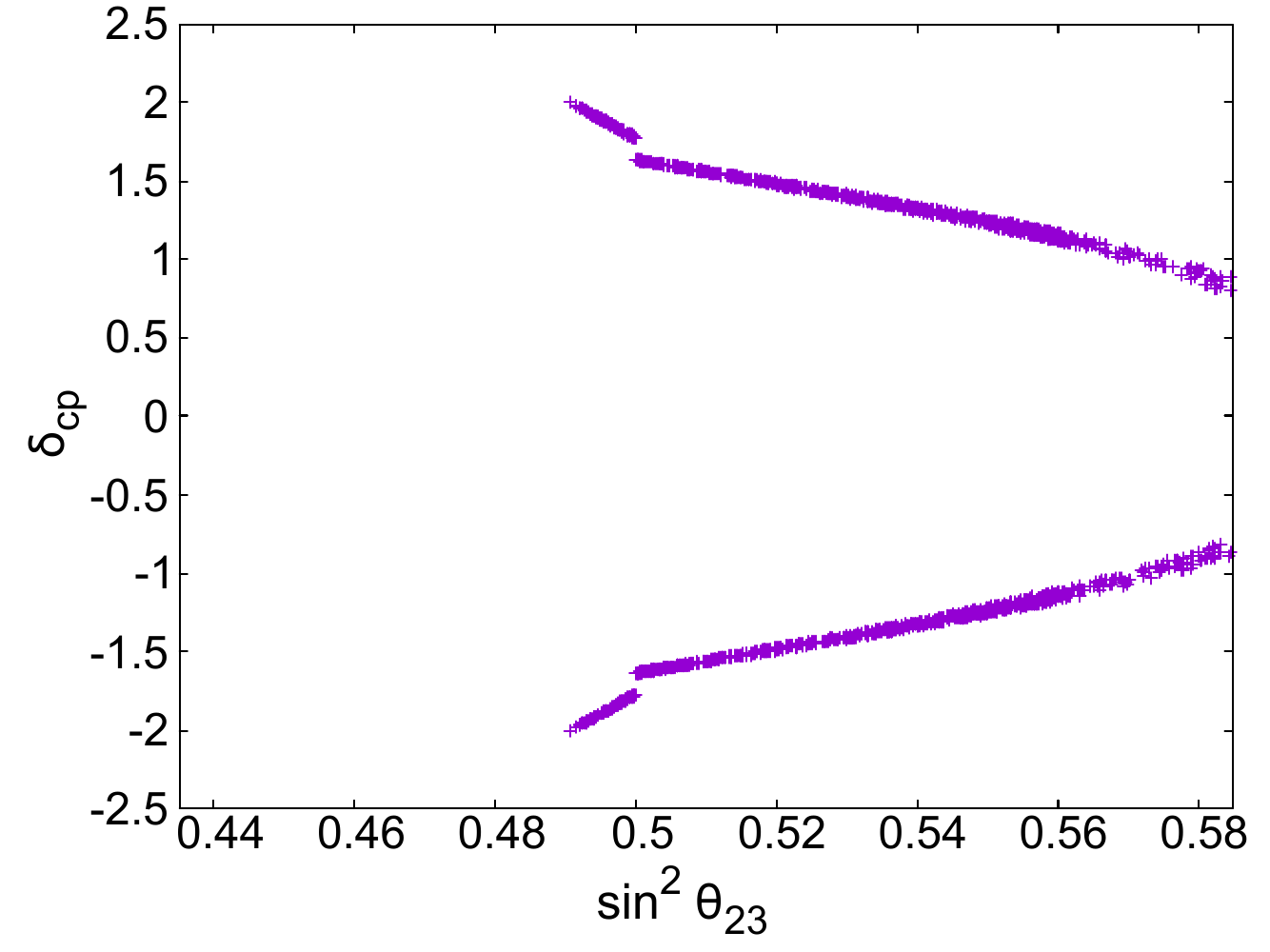}
        \subcaption{The $\sin \theta _{23}$-$\delta _{CP}$ in the NH case.}
    \end{minipage}
    \begin{minipage}[b]{0.48\linewidth}
        \includegraphics[width=0.9\columnwidth]{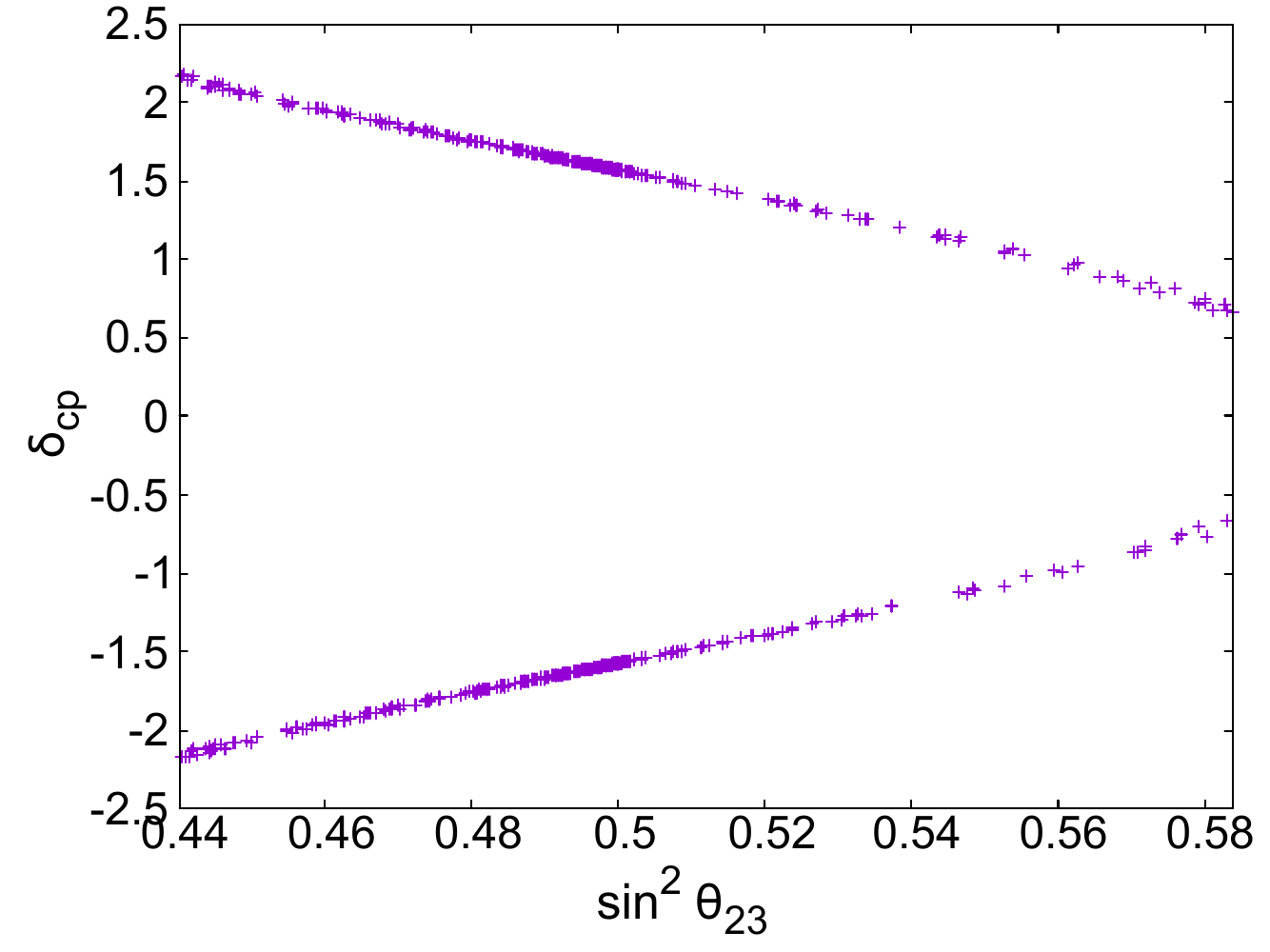}
        \subcaption{The $\sin \theta _{23}$-$\delta _{CP}$ in the IH case.}
    \end{minipage}
    \caption{The allowed regions of the lepton mixing anlge and Dirac CP phase.}
\label{fig:mixing-angle-and-delta-CP}
\end{figure}

In our numerical analyses, we use the NuFit-6.0 data~\cite{Esteban:2024eli}. 
The allowed regions for mixing angles and CP violating phases are same in Ref.~\cite{Morozumi:2017rrg}. 
Hereafter we consider the special case in the right-handed Majorana neutrinos in Eq.~\eqref{eq:right-handed-Majorana-neutrino-mass-matrix} such that 
we assume the only $y_\xi $ is complex parameter, while the $y_{\xi '}$ is assumed a real parameter, for simplicity\footnote{We can take the coupling $y_{\phi _S}$ for Eq.~\eqref{eq:right-handed-Majorana-neutrino-mass-matrix} as real parameter in general.}. 
In Fig.~\ref{fig:mixing-angles}, we show the allowed region for the $\sin^2 \theta_{12}$ and $\sin^2 \theta_{13}$ within 3$\sigma$ standard deviation for NuFit-6.0 data~\cite{Esteban:2024eli}.
We also show the allowed regions for the $\sin ^2\theta _{23}$ and 
Dirac CP violating phase $\delta _{CP}$ in the normal hierarchy (NH) and inverted hierarchy (IH) case which are shown in  Figs.~\ref{fig:mixing-angle-and-delta-CP} (a) and (b), respectively. 
Here we take into account for the CP asymmetry numerically, which are realized the BAU of the present universe, since there are rich complex parameters in Eq.~\eqref{eq:Dirac-Yukawa-matrix-NL}.  
The $\sin^2 \theta_{23}$ and the CP violating phase $\delta _{CP}$ for the NH and IH case are shown in Fig.~\ref{fig:mixing-angle-and-delta-CP}(a) and (b), respectively. 
In the IH case, the allowed regions are same as Ref.~\cite{Morozumi:2017rrg} even if we consider the special case such that 
we assume the only $y_\xi $ is complex parameter, while the $y_{\xi '}$ is assumed a real parameter in Eq.~\eqref{eq:right-handed-Majorana-neutrino-mass-matrix}.

\begin{figure}[h]
  \centering
    \begin{minipage}[b]{0.48\linewidth}
        \includegraphics[width=0.9\columnwidth]{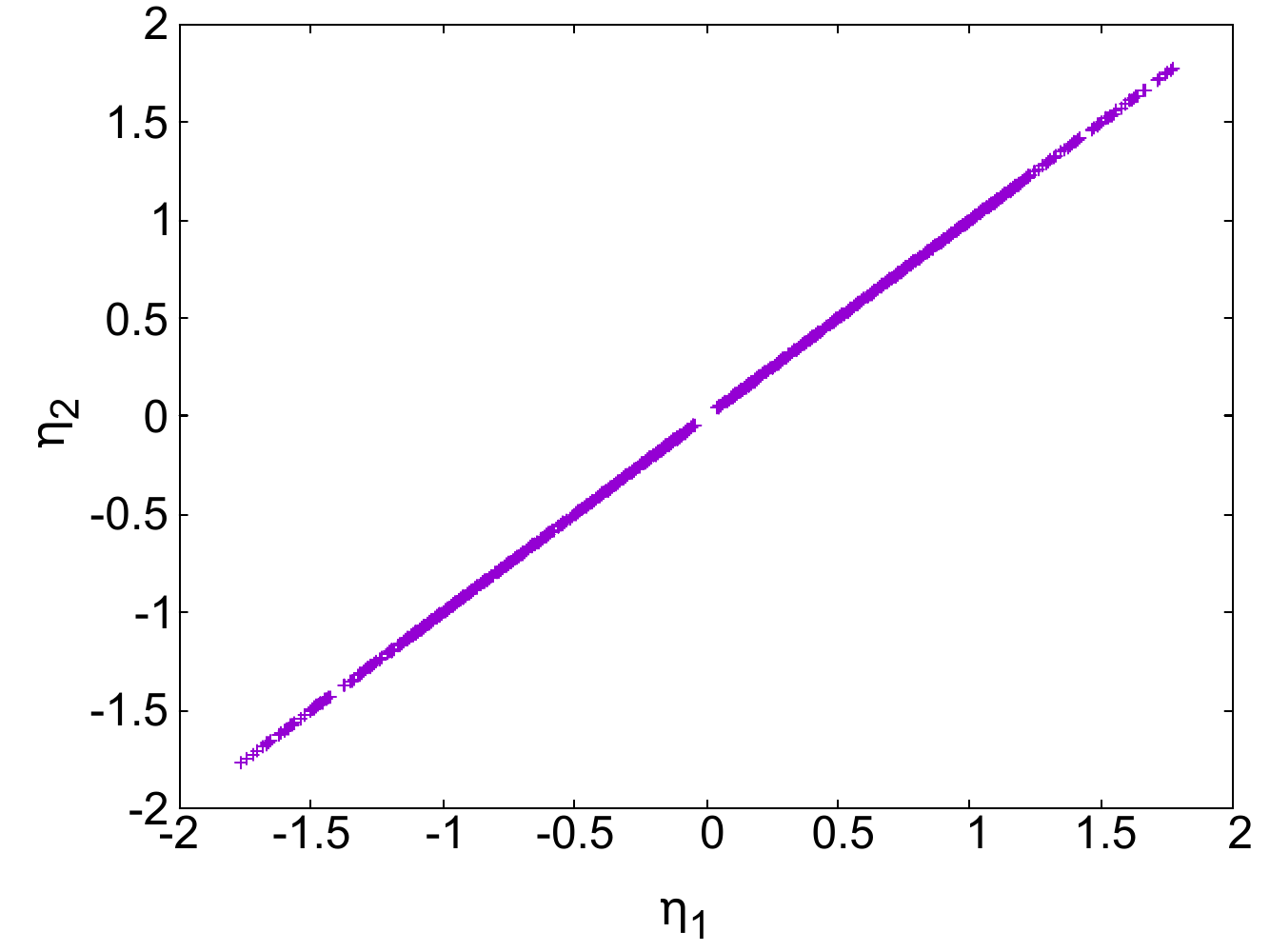}
        \subcaption{The $\eta _1$-$\eta _2$ plane in the NH.}
    \end{minipage}
    \begin{minipage}[b]{0.48\linewidth}
        \includegraphics[width=0.9\columnwidth]{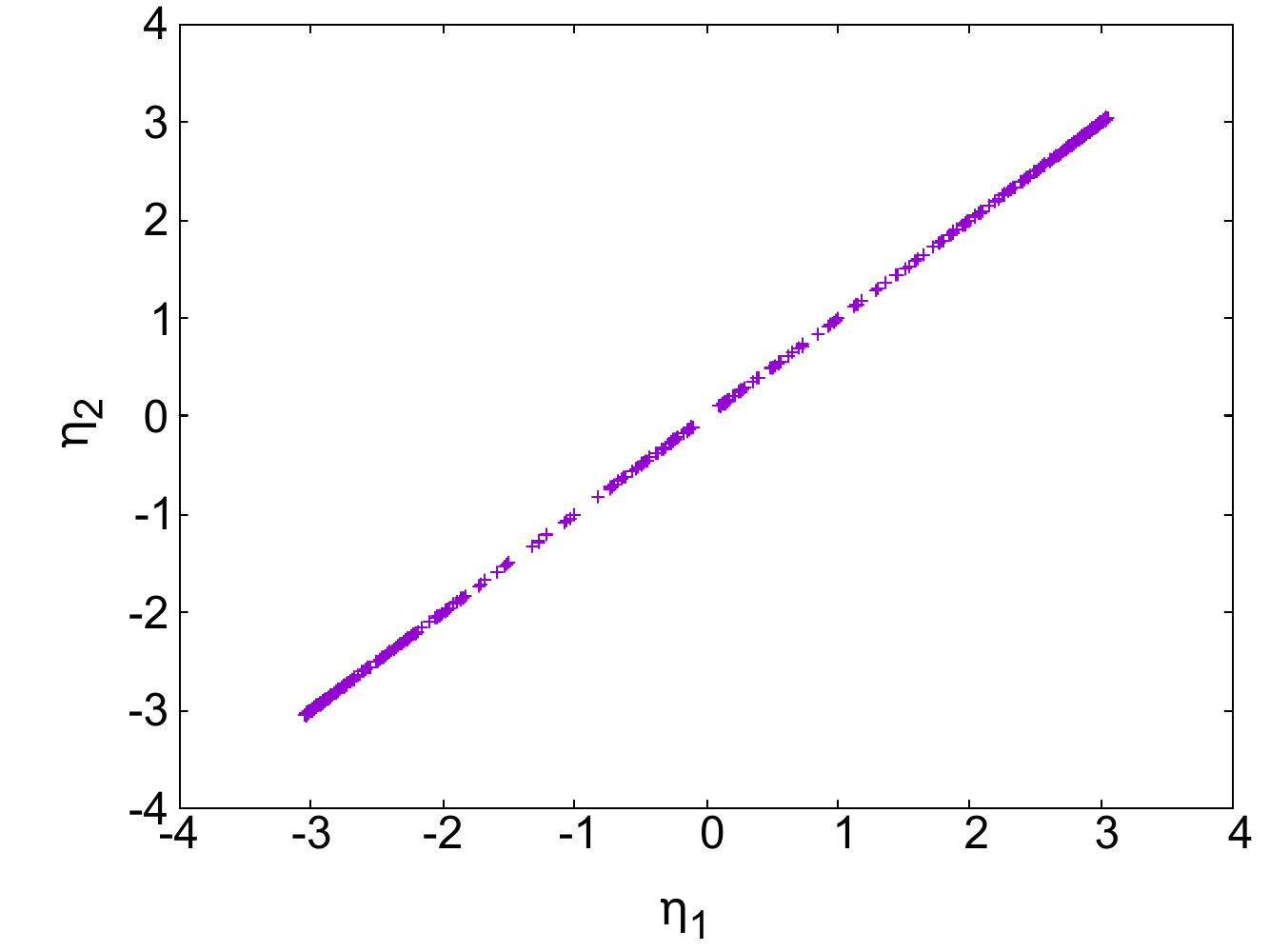}
        \subcaption{The $\eta _1$-$\eta _2$ plane in the IH.}
    \end{minipage}
    \caption{Majorana phases.}
    \label{fig:Majorana-phases}
\end{figure}

The Majorana phases $\eta_1$ and $\eta_2$ are determined by using PMNS matrix as follows:
\begin{align}
  \eta_1 = arg \left[\frac{U_{e1}^{mod} U_{e3}^{mod *}}{c_{12}c_{13}s_{13}e^{i \delta_{CP}}} \right],\qquad 
  \eta_2 = arg \left[\frac{U_{e2}^{mod} U_{e3}^{mod *}}{s_{12}c_{13}s_{13}e^{i \delta_{CP}}} \right].
\end{align}
In Figs.~\ref{fig:Majorana-phases}(a) and (b), we can predict the Majorana phases $\eta _1$ and $\eta _2$ for the NH and IH, respectively. 
By using mixing angles $\theta _{ij}$, CP violating phase $\delta _{CP}$, 
we consider the effective mass for the 0$\nu \beta \beta$ decay.
The effective mass for the $0\nu \beta \beta$ decay is determined by the magnitude of the lightest neutrino mass, the other neutrino mass eigenvalues, and the Dirac and Majorana CP violating phases. 
The effective mass $|m_{ee}|$ for the $0\nu \beta \beta$ decay is expressed as follows:
\begin{align}
|m_{ee}| &= \left| \sum_{i = 1}^3 m_i U_{e i}^{mod~2} \right|\\
 &= \left| m_1 U_{e1}^{mod~2} + m_2 U_{e2}^{mod~2} + m_3 U_{e3}^{mod~2} \right|
\end{align} 
\begin{figure}[h]
\centering
   \begin{minipage}[b]{0.48\linewidth}
        \includegraphics[width=0.9\columnwidth]{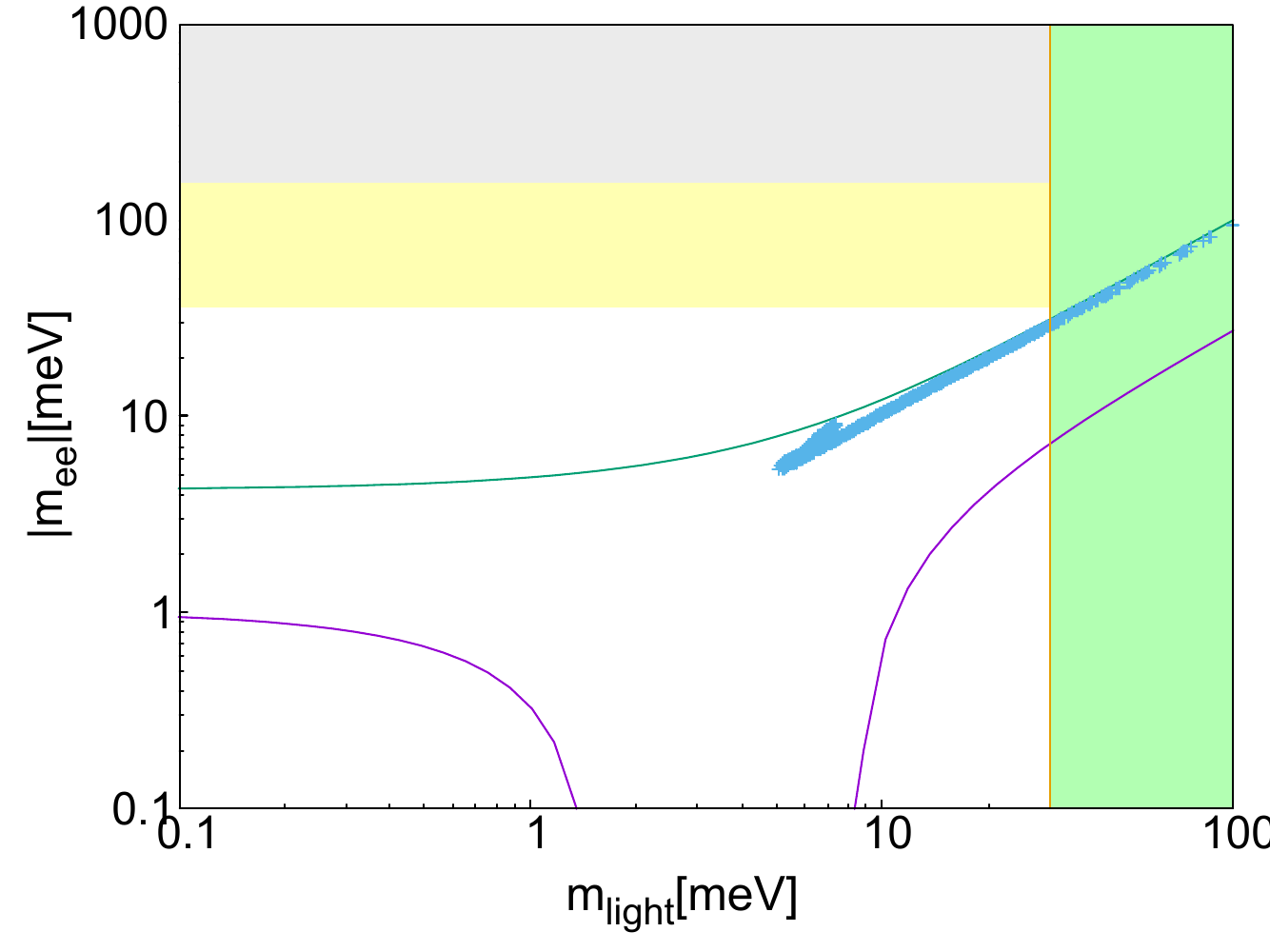}
        \subcaption{The $m_\text{light}$-$|m_{ee}|$ plane in the NH.}
    \end{minipage}
    \begin{minipage}[b]{0.48\linewidth}
       \includegraphics[width=0.9\columnwidth]{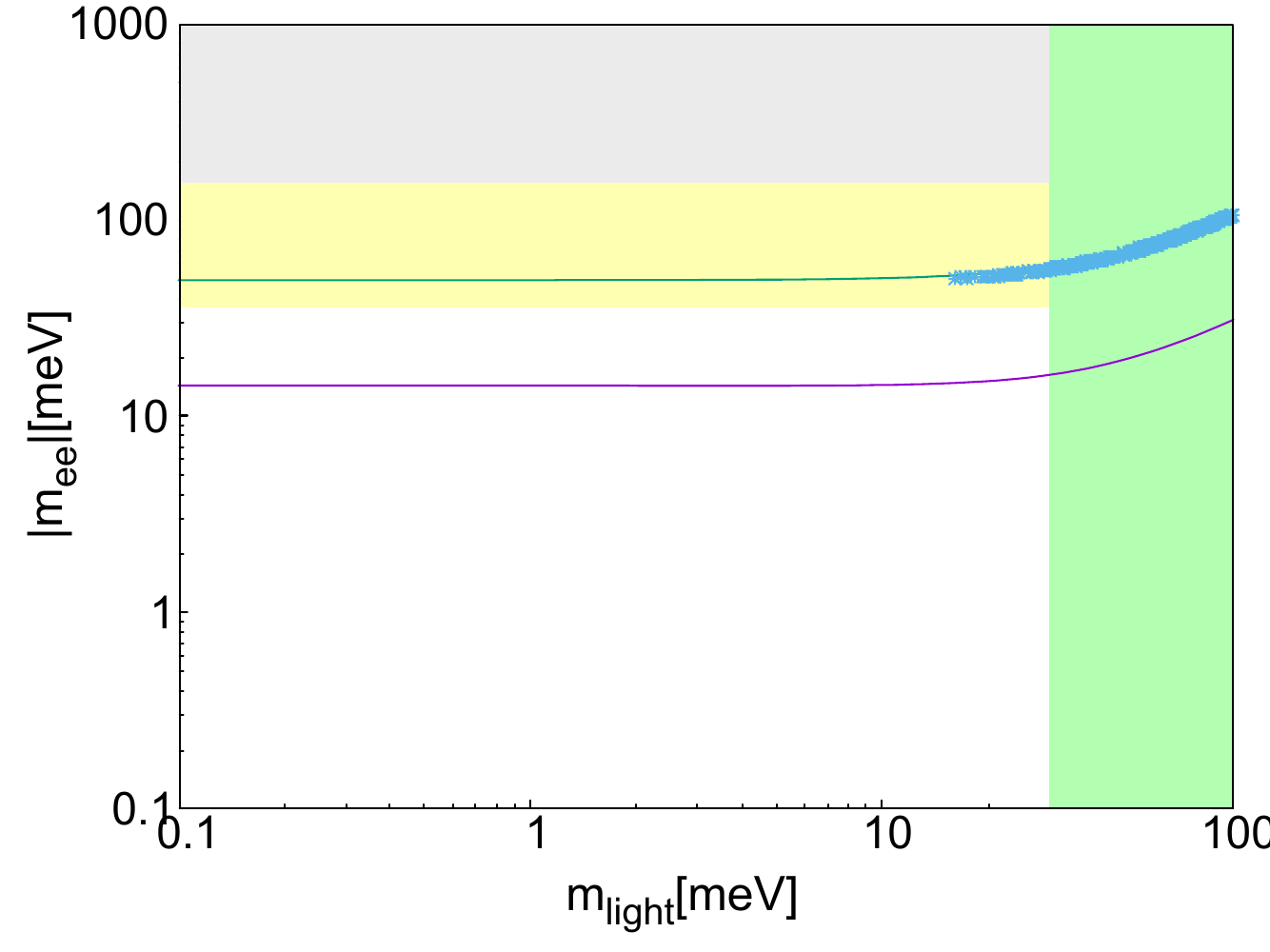}
       \subcaption{The $m_\text{light}$-$|m_{ee}|$ plane in the IH.}
    \end{minipage}
   \caption{The effective mass for the 0$\nu \beta \beta$ decay.}
   \label{fig:mlight-0nubetabeta}
\end{figure}
In Figs.~\ref{fig:mlight-0nubetabeta} (a) and (b), we show the lightest neutrino mass and effective mass for the $0\nu \beta \beta $ decay. Since we assume that one of the couplings for the right-handed Majorana neutrinos is only complex parameter $y_\xi $, while the others are real parameters $y_{\Phi _S}$ and $y_{\xi }$, we can predict narrow regions.
In Figs.~\ref{fig:mlight-0nubetabeta}, the green region is the upper limit on the lightest neutrino mass of $30.1$ meV at 90 \% C.L.\cite{kamland},
the gray region is excluded, and the yellow region is the upper limit on the effective Majorana neutrino mass of 30.1 -- 136 meV as estimated in Ref.~\cite{Planck:2018vyg}.
We find that the lightest neutrino mass is at least $5$ meV and $15$ meV for the NH and IH, respectively. 

Next we discuss the diagonalization for the right-handed Majorana neutrinos in order to obtain the CP asymmetry in Eq.~\eqref{eq:Dirac-Yukawa-matrix-NL}. 
We can diagonalize the $M_{\nu} M_\nu ^\dagger$ by using the unitary matrix $U_\nu$.
We can also diagonalize the complex symmetric matrix $M_\nu$ by using $U_\nu$ as follows:
\begin{align}
  U_R^\dagger  M_R U_R^* = diag(M_{R1} e^{i \phi_{R1}}, M_{R2} e^{i \phi_{R2}}, M_{R3} e^{i \phi_{R3}}).
\end{align}
In order to remove these phases, multiply phase diagonal matrix $P_R$ = diag$(e^{i\frac{\phi_1}{2}}, e^{i\frac{\phi_2}{2}}, e^{i\frac{\phi_3}{2}})$ on both sides:
\begin{align}
  M_R^{diag} &= P_R^\dagger \left(U_R^\dagger  M_R U_R^* \right) P_R^*\notag\\
  & = diag(M_{R1}, M_{R2}, M_{R3}).
\end{align}
Then, the unitary matrix $U_R P_R$ makes the mass matrix real diagonal.

Finally, we comment on the BAU at the present universe. In our calculations, we can realize the BAU at the present universe, since we have rich free complex parameters in Eq.~\eqref{eq:Dirac-Yukawa-matrix-NL}.

\section{Summary}
\label{sec:Summary}
In this paper, we revisited the supersymmetric flavor model based on the symmetries for the $SU(2)_L \times A_4 \times Z_3 \times U(1)_R$, which extends the original AF construction by incorporating flavon and driving superfields in oreder to obtain the non-zero reactor angle. 
The flavon VEVs are aligned through the minimization of the scalar potential, leading to a predictive structure for the charged lepton and neutrino mass matrices. 
We found that the model successfully reproduces the observed lepton mixing angles and mass squared differences within the current experimental bounds.

Furthermore, we explored the possibility of generating the observed BAU via thermal leptogenesis in this framework. 
The model introduces three right-handed Majorana neutrinos, whose masses and Yukawa couplings are controlled by the flavor symmetry. 
At leading order, the model yields a simple Dirac neutrino Yukawa structure which, although consistent with the type-I seesaw mechanism, fails to generate sufficient CP asymmetry required for successful leptogenesis. 
To address this issue, we incorporated NLO corrections in the Dirac neutrino sector. 
These NLO terms introduce complex parameters and perturb the original texture in a controlled manner, thereby enabling non vanishing CP violating effects essential for generating a lepton asymmetry via the decay of the heavy right-handed Majorana neutrinos. 

Through a numerical scan over the model parameters, we identify viable regions in which both the neutrino oscillation data and the observed baryon asymmetry can be simultaneously explained. In particular, we found that for both normal and inverted ordering of neutrino masses, the model can accommodate successful leptogenesis considering the NLO for our model. Then, we can realize the BAU at the present universe. 
We also found that the lightest neutrino mass is at least $5$~meV for the NH, and $15$~meV for the IH, respectively. 
Correspondingly, the effective Majorana mass lies within the sensitivity reach of upcoming the $0\nu\beta\beta$ experiments.
The range of the Majorana phases may be tested in future experiments. 

This study highlights the interplay between flavor symmetries, CP violation, and cosmological baryogenesis. 
The constrained structure of the model offers testable predictions that can be confronted with future experiments in both neutrino physics and cosmology. 
Continued improvements in the measurements of the Dirac CP phase, the neutrino mass ordering, and the $0\nu\beta\beta$ decay rate will provide important opportunities to further scrutinize the viability of this framework. 
As such, this model represents a concrete and predictive realization of flavor driven leptogenesis.


\vspace{1cm}
\noindent
{\large \bf Acknowledgement}
\vspace{1mm}

The work of T.~N. was supported by the Fundamental Research Funds for the Central Universities.
The work of T.~T. was supported by JST SPRING, Grant Number JPMJSP2121.


\appendix
\section*{Appendix}

\section{Multiplication rule of $A_4$ group}
\label{sec:multiplication-rule}
We use the multiplication rule of the $A_4$ triplet as follows:
\begin{align}
\label{eq:multiplication-rule}
\begin{pmatrix}
a_1\\
a_2\\
a_3
\end{pmatrix}_{\bf 3}
\otimes
\begin{pmatrix}
b_1\\
b_2\\
b_3
\end{pmatrix}_{\bf 3}
&=\left (a_1b_1+a_2b_3+a_3b_2\right )_{\bf 1}
\oplus \left (a_3b_3+a_1b_2+a_2b_1\right )_{{\bf 1}'} \nonumber \\
& \oplus \left (a_2b_2+a_1b_3+a_3b_1\right )_{{\bf 1}''} \nonumber \\
&\oplus \frac13
\begin{pmatrix}
2a_1b_1-a_2b_3-a_3b_2 \\
2a_3b_3-a_1b_2-a_2b_1 \\
2a_2b_2-a_3b_1-a_1b_3
\end{pmatrix}_{{\bf 3}}
\oplus \frac12
\begin{pmatrix}
a_2b_3-a_3b_2 \\
a_1b_2-a_2b_1 \\
a_3b_1-a_1b_3
\end{pmatrix}_{{\bf 3}\  } \ .
\end{align}
More details are shown in the review~\cite{Ishimori:2010au,Ishimori:2012zz}.


\end{document}